\documentclass[UTF8,a4pape,]{article}

\usepackage{amsmath,amsthm,verbatim,amssymb,amsfonts,amscd, graphicx}
\usepackage{graphics}
\theoremstyle{plain}
\usepackage{amsmath} 
\usepackage{braket}

\begin{document}
\title{\bf Multispin interaction of plaquette lattice in SU(N) system}
\author{Chenhuan Wu
\thanks{chenhuanwu1@gmail.com}
\\Deparment of physics, Northwest Normal University, China}

\maketitle
\vspace{-30pt}

\begin{abstract}
\begin{large}
This article consider a situation of SU(N) system with broken symmetry, and therefore the spin-liquid phase is exist in the phase transition stage.
And explore the antiferromagnetic spin interaction with long range order in a two dimension square lattice using the two alkaline earth atoms
state short range interaction.
The appearance of spin liquid due to noncolline mechanism which causing by destruction of long range order in this model is mentioned.
In a N-site fermion-model system, 
I analyse the effect of fluctuation on order and phase transition. As well as this N-component spin system with disordered spin, the critical phenomenon is showed to
reflect the fluctuation effection on process of phase transtion. 
\end {large}
\end{abstract}

\begin{large}
Recently there are some experiments about the interaction between the atoms widely used for experimentation like the alkaline-earth atoms and alkali atoms,
and further indicate the filling fraction $1/\tau$ and the number of spin component $N$ is closely connected to the system stability under the influence of fluctuation.
The hydrogen and alkaline-earth atoms are most widely used in the preparation hyperfine structure of ground state
because the hyperfine splitting of ground state of this kind of atoms (about 1 GHz) bigger than orther atoms and that's beneficial for 
seeking the appropriate frequency's laser to prepare the desired ground state,and that works in the case of excited
state in some case.
Quantized vortices and fermion pairing(the Cooper pairs) are important phenomenons in superfluidity
(like liquid $^{3}$He or $^{4}$He) and superconductivity, to investigate it,a experiment \cite{Zwierlein M W}of 
BEC-BCS crossover presenced by a Feshbach resonance,and the Fermi gas expansioned by reduce the laser power. Another one\cite{Wille E} of
the resonance in $l=1$ ( $l$, angular momentum, i.e.,the p-wave) was observed in compressed Fermions which by increasing 
the power of optical trap. For the quantized vortex of such a topological excitation, the significant U(1) 
symmetry broken in a normal superconductor pahse transition have showed in the first experiment in the above,
it's a process of increasing of couping strength between two Fermions,the couping strength proportional to interaction strength
 which introduced by parameter $1/k_{F}a$ (a,sattering length;$k_{F}$,Boltzmann constant). Ref.\cite{Machida M} treat these two side of process as the 
conserved and damping pair represent the boson condensation and Fermion pairs condensation respectively.

Superconductivity under BCS Framework is always the BEC,
just like the fact that we can't observe a condensate phenomenon of electron pairs in the independent atoms.
since each experiment practiced by control the magnetic filed, the produced Aharonov-Bohm phase in boson condencetion
general the Meissner effect and the goldstone model was eliminated by Anderson-Higgs mechanism.the same symmetry broken
in BEC-side is the degree of freedom of phase disappears due to the conjugation relation between phase and total number of particles,
. While the fluctuates of number of particles enhance (required by off-diagonal long-range order (ODLRO)),the
 phase fluctuation is reduced by particle-interaction and therefore the gauge sysmetry is broke ,the SU(N)-symmetry-broken
which we will discuss later,and condense occur,here the
condensation can be intuitively seen in the single particle reduction density matrix of the Bose gas system.

No matter the experiment of interaction between halfly spin mixture of lowest hyperfine states of $^{6}Li$\cite{Zwierlein M W},
or just between atoms populate in ground state and excited state with same spin, both show that the interaction
is proportional to the scattering length $a$, e.g.,in the BEC-BCS crossover the Fermions interaction (not the coupling) 
decreases with the increasing scattering length between the two state of fermionics.the interaction parameter $1/k_{F}a$ appear
minus in the BCS-side\cite{Zwierlein M W}.

Here the fermionics in these experiment is
make full use of the excellent propertices of alkali-like atoms,for example, $^{6}Li$ ,$^{40}K$ and $^{87}Rb$. 
Here we can obtain the densities of momentum distribution by means of the time-of-flight image ,which realized by the way of optical absorption, has been showed
in the ref.\cite{Taglieber M,Anderlini M}. Note that the applied magnetic field switch during this time. these distuibution are also important 
in the investigate of energy scale changes or phase transitions in the two-atom systems, which also reflected by the double-well phase digram.
 
For SU(N) system, in the case of low-N which corresponding to low symmetry, it contains not only the degrees of
spin, but the different energy level of atoms (i.e., the electron orbital degrees of freedom). That rise up the dependence between unclear spin and scatteing, 
For a mixture system mixing by same-orbit atoms interaction and different-orbit atoms interaction of Hubbard model,
as show in the Fig.2,which we use the excellent typical cuprate phase diagram\cite{Ghaemi P,Gorshkov A V}.

For the SU(N) system formed by the alkaline earth atoms which N is higher, means that it's more easy to realize the high degree of symmetry due to the missing of strong interaction 
from hyperfine struction which appear in alkali atoms. And that breaking the strong coupling between nuclear spin and electronic degrees of freedom. Denotes 1 and 2 are the two 
states which the two sublattices $i$ and $j$ in respectively(see fig.1). The fig.1a shows the coupling strength of nearest-neighbour and next nearest-neighbour 
sublattices, and fig.1b shows the SU(N) antiferromagnets in a four singlet plaquette lattice. 
                                                          The $i$ and $j$ contains alkaline-earth atoms in different
Zeeman state in a two-dimension square lattice show in Fig.1, which describe the square lattice consisted by singlets(circle in the dotted line oval inside) and triplets.
the lattice remain the AF spin long range order
the coupling strength J also reflect the energy of isotropic tunneling\cite{Gorshkov A V}.In this model, the potential well contain atoms of ground state and excited
state wound be fabricted by using a appropriate radio frequency (RF) to remove the potential barrier and merge the single-particle potential wells.this way is also
suitable in quantum qubit controling\cite{Anderlini M}. In general,we can use the laser to manipulate the potentialand therefore control the lattice topology.
Note that although the atoms collision loss obey the high symmetry due to large-N, it's still controlled
indirectly by unclear spin,like the repression effect from Fermion statistic. 

\begin{figure}[!ht]
   \centering
   \begin{center}
     \includegraphics*[width=0.7\linewidth]{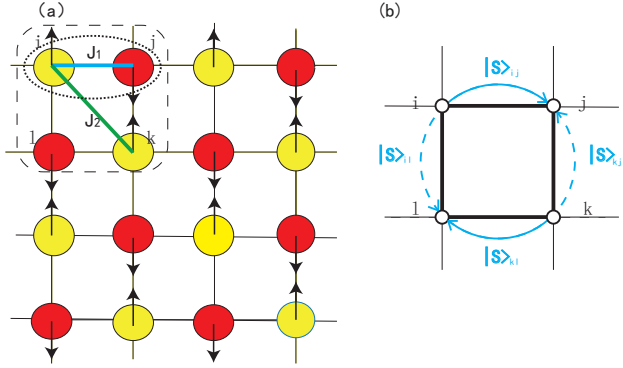}
   \caption{The model of square lattice which contain of many four-bond plaquette lattices. The yellow and red spheres represent the sublattice $i$ and $j$, respectively. The upward-arrow and downward-arrow 
instructions the total spin direction.the dotted line
oval marked out a antiparallel-spin singlet which constrainted in one potential well. The blue line and green line indicate the coupling between the two sublattice
 in a singlet and coupling between the next 
neighbors with same spin direction,respectively.And we use $J_{1}$ and $J_{2}$ to denote these two coupling: the $J_{1}$ is the coup between the nearest-neighbour 
sublattices and $J_{2}$ is the next nearest-neighbour one. The four site $ijkl$ occupied by four sublattices and form a plaquette lattice.
(b) shows the four singlets in a plaquette, which correspond to the square dashed box in (a).}
   \label{fig:user-item-bipartite}
   \end{center}
\end{figure}

\begin{figure}[!ht]
   \centering
   \begin{center}
     \includegraphics*[width=0.8\linewidth]{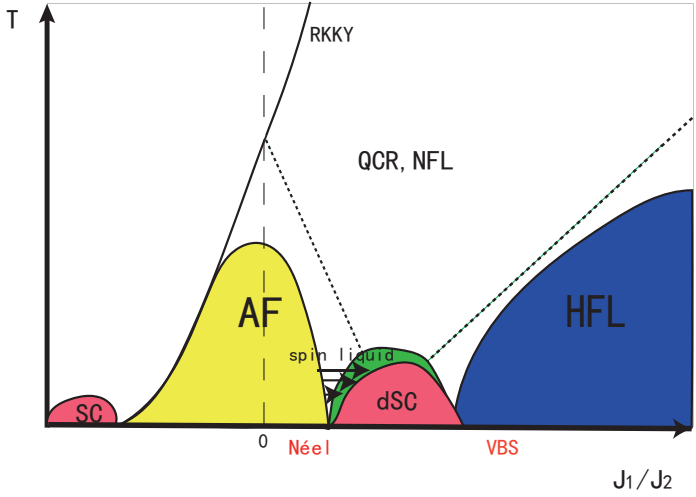}
   \caption{The T phase diagram of versus $J_{1}/J_{2}$ of cuprate.the vertical dash populate in limit of $J_{2}\geq J_{1}$ cross the most part of antiferromagnetic (AF) Mott insulator. 
The red area of SC is
superconductor and dSC is d-wave superconductor(due to the high $H_{T}c$uprate superconductors are consisting of d-wave pairs\cite{Ghaemi P}).
And the heavy-Fermi liquid (HFL) which suggested due to the low-temperature Kondo excitation\cite{Coleman P} appear in $J_{2}\geq J_{1}$ area.
the green area above the dSC is the nernst region
which as a transition zone to the superconductor.the RKKY-line is the Ruderman-Kittel-Kasuya-Yosida interaction between the two atoms in a triplet which mediated
by the singlet one,it's one kind of long-range interaction here, the line rise up obviously with the increasing $J_{1}/J_{2}$. The sector area between the AF and HFL is the quantum critical
region(QCR), which is limited by two dot lines,and one of its characteristic is the non-Fermi liquid. there is a quantum critical point exist at the intersection of two dot lines.
 The black arrows which point to the dSC part denote the critical spin-liquid under zero temperature 
caused by the 
noncollinear N$\acute{e}$el.In addition,there is pseudo spin gap appear between AF and dSC,we didn't show this in the phase diagram since it's unstable.}
   \label{fig:user-item-bipartite}
   \end{center}
\end{figure}

The Hamiltonian as below\cite{Gorshkov A V},
\begin{equation}   
\begin{aligned}
H=&\sum_{\langle i,j\rangle,\alpha}[a^{ij}_{\alpha}n_{i\alpha}n_{j\alpha}+b^{ij}_{\alpha}S^{n}_{m}(i,\alpha)S^{m}_{n}(j,\alpha)]\\
&+\sum_{\langle i,j\rangle}[a^{ij}_{12} n_{i1} n_{j2}+a^{ji}_{12} n_{j1} n_{i2}+b^{ij}_{12} S^{n}_{m}(i,1) S^{m}_{n}(j,2)+b^{ji}_{12}S^{n}_{m}(j,1)S^{m}_{n}(i,2)\\
&+c^{ij}_{12}S^{2m}_{1m}(i)S^{1n}_{2n}(j)+c^{ji}_{12}S^{2m}_{1m}(j)S^{1n}_{2n}(i)+d^{ij}_{12}S^{2n}_{1m}(i)S^{1m}_{2n}(j)+d^{ji}_{12}S^{2n}_{1m}(j)S^{1m}_{2n}(i)]
\end{aligned}
\end{equation}

($\alpha$,electron state;a,stands nearest-neighbour interaction;b,c and d,stands nearest-neighbour exchange of spin, pseudo-spin, atomic state,respectively:
n,atom number; m, Zeeman energy levels)
the permutation operator contained in later terms $S^{n}_{m}(i,\alpha)=c^{\dag}_{i\alpha m}c_{i\alpha n}$, $S^{m}_{n}(j,\alpha)=c^{\dag}_{j\alpha n}c_{i\alpha m}$,
where $\langle i,j\rangle$ denote a nearest pair of lattice sites, $c^{\dag}$ and $c$ are creation operator and annihilate operator take effect on designated sites, respectively.
In this model, for SU(N) antiferromagnets(singlet) ,its energy lower than the triangle one,i.e.,$E_{S}<E_{T}$,
due to the more strong confinement origin from the asymmetry swap gate voltage in quantum computation. 
The much powerful $J_{1}$-producing ionization is useful.the rare-earth atoms is cyclopedic 
used for the better confinement effect due to its large spin degeneracy. Fig.2 show the ratio of coupling $J_{1}/J_{2}$ show the transition of N\'eel order to VBS (valence bond solid) order.
this transition has been implement by first order transition\cite{Tissier M} or second order transition\cite{Ghaemi P}, and the critical value of coupling has been
calculated using the histogram quantum Monte Carlo (QMC) method\cite{Kaul R K} and theoretical analysis\cite{Ghaemi P}. Indeed, it's a eliminate process of the singlet
which pairly correlate in long range, and translate to valence bond singlet (or triplet) in a ladder model or plaquette model and so on, a summarization of 
 dimerized models is presented in Ref.\cite{Bogacz L}.

In fig.2 it's easy to see that 
as the N increase (i.e., along the positive part of the horizontal axis), the symmetry increase and result in the decouping between unclear spin and
degree of freedom of electrons. With the $J_{1}/J_{2}$ grow up, we see that accompanied by
the $J_{2}$ decrease and $J_{1}$ increase, i.e.,the coupling between singlet coupled more closely and more weaken for triplet, and also, 
the state energy $E_{T}$ decrease
, $E_{S}$ increase, the triplet
particle-hole motion destroy the AF, and causing the broken of magnetic long-range order with the growing temperature higher than $T_{c}$ though still remain some
long-range magnetic order in microscopic scale\cite{Qin H J}. Note that the part of the right side of the zero axis is the spin singlet for the overall phase,no matter
the N$\acute{e}$el phase or the VBS phase and the quantum dimer phase as mentioned below. 
Here the broken of long-range order can be related to the corresponding classical case of Ising model,
as N increase, the large number of sites result in more larger quantum fluctuation, so that the long-range order of spin, orbit, and magnetization are destroyed.
It's also appear in the fermi face as a phenomenon of "magnetic breakdown"\cite{Weng H}.
in higher dimension model, the more degrees of freedom is helpful in preserve the order parameters, 
that's why the N\'eel phase is not exist in one-dimensional spin model. But it's possible for a orderly state formed in the disorder state is exist due to the 
quantum fluctuations state may choice a subset of multiple degenerate state and construct a new spin order\cite{Villain J}.

Furthermore, the same as the elastic Ising model(EIM), the interaction effect $J_{12}$ is particle particle spacing $\bf r$-dependent, resulting to the scattering length-dependent 
interaction parameter $C=4\pi h^{2}a_{s}/m$, m is mass of atoms, and $a_{s}$ is the s-wave scattering length. The interaction effect $J_{12}$ can be written

\begin{equation}   
\begin{aligned}
J_{12}=\int w^{*}(r-r_{i})H_{12}w(r-r_{j}) d^{3}r
\end{aligned}
\end{equation}

Where $w$ is the wannier function and $H_{12}$ is the interaction Hamiltonian as $H_{12}=-\frac{\hbar^{2}}{2m}\triangledown^{2}+U_{12}(r)$,$U_{12}$ is interaction energy.
we also show the spin liquid in the figure. In fact, we can find the critical spin liquid in two dimension or three dimension under 
low temperature $T<T_{c}$ and $N_{1}=N_{2}=2$ (for the sublattice 1 and 2) in two dimension square lattice for SU(N) system\cite{Gorshkov A V},
 it has been observed in $LiNiO_{2}$ with with short-range ferromagnetic order\cite{Kitaoka Y}. These spin liquid can be utilized 
in the preparation of Mott insulator\cite{Ghaemi P} (the effection of Mott insulator increases as the gap of spin liquid increases) 
and mapping out the superfluid regimes in a absorption image\cite{Zwierlein M W}.
                        
The vertical dashed line which indicate the $J_{2}/J_{1}$
represent the $J_{2}=0$-limit of VBS-side. For the simplified system of two-dimension square lattice \cite{Gorshkov A V,Kaul R K},the part with $N>4$ 
is the VBS-order state and $N=2,3,4$
corresponding to N\'eel-order state,and that's also the range where N\'eel-VBS transition happen.
That means the critical value of N is close to 4 (4.57) in two dimension square lattice\cite{Beach K S D}, 
and a lot of research have proved that this critical value is close to 5 (5.1) for three-dimension stacked triangular lattice\cite{Beach K S D,Pelissetto A}.

So here the Hubbard Hamiltonian is,

\begin{equation}   
\begin{aligned}
H=\sum_{\langle i,j \rangle,m}[J_{1}(S^{j}_{i}(1,m)S^{j}_{i}(2,m)+h.c.)]+\sum_{\langle i,k \rangle,m}[J_{2}(S^{j}_{i}(1,m)S^{j}_{i}(2,m)+h.c.)]+V_{ex}S^{m'}_{m}S^{m}_{m'},
\end{aligned}
\end{equation}

It's a result of consider both the spin and orbital degrees of freedom, where $m$ and $m'$ repersent different states under atoms exchange interaction. Note that the above equation are fit the situation of large N which the high symmetry
guarantee the indepence between unclear degree of freedom and electrons. in the case of low symmetry (N is small), then atoms may occupy the other states (gapless 
Brillouin zones in optical well manipulation) which not just
m and $m'$. so we will need to add it into the exchange term in equ(3). Here the interaction energy $U_{12}=2C\int|\phi_{1}(x)|^{2}|\phi_{2}(x)|^{2}d^{3}x$.

I denote the states triplet as $|T\rangle$ and contains 1,0,-1 and singlet as $|S\rangle$, and defines Hubbard operator $X^{11}=|1\rangle\langle 1|$. For  
completeness, it have $X^{SS}+X^{11}+X^{00}+X^{-1-1}=0$, the term $X^{SS}$ is the local singlet projection which has $X^{SS}=-\frac{1}{L^{d}}S^{n}_{m}(i)S^{m}_{n}(j)$ for
singlet $|S\rangle_{ij}$.
 Using the permutation operator, it also have 
$S^{m}_{m}/2=X^{11}$, $S^{m+1}_{m+1}/2=X^{-1-1}$, $S^{m+1}_{m}/\sqrt{2}=X^{10}+X^{0-1}$, $S^{m}_{m+1}/\sqrt{2}=X^{01}+X^{-10}$, where m are the Zeeman 
states correspond to unclear spins. These four formulas descript the interaction between the three state within a triplet state, and I denote it as
 $|T\rangle$.
Additionally, I define the interaction between state of singlet and triplet as $|ST\rangle$, it can be obtain that $|ST\rangle^{+}=\sqrt{2}(X^{1S}-X^{S-1})$,
$|ST\rangle^{-}=\sqrt{2}(X^{S1}-X^{-1S})$,and $|ST\rangle^{z}=-X^{0S}-X^{S0}$. here $|ST\rangle^{z}$ is the component of total interaction $|ST\rangle$ which 
rotate around the $z$ axis for $\pi$\cite{Melko R G}.
and it's easy to find that $|T\rangle$ and $|ST\rangle$ form a set of vertical orthogonal basis in the ground state and therefore there is exist a parameter space 
which have SU(N) symmetry. There are also some other ways to show the symmetry of particles behavior, like the Kitaev model\cite{Baskaran G} which is important in 
quantum computation and the 
Gamma matrix model which has a global Ising symmetry in a plaquette\cite{Yao H}.

In this plaquette lattice model. Similarity, for the triangular lattices sphere, we can also obtain the SO(3) symmetry 
which extended to three-dimensional structure by piling the planar elementary cell in a lattice through the similar way of construction\cite{Peles A,Delamotte B}.

A more universally connection have been given by K.S.D.Beach $et\ al.$ in ref.\cite{Beach K S D}. But a insufficiency is that there method is nonorthogonal and consider 
only one column of Young tableau, i.e., consider only one orbit state and ignore the degree of freedom of orbits.
The $|T\rangle$ and $|ST\rangle$ here are both a three component state vector, and solve this insufficiency here with a global symmetry.

As shown in the fig.1b, the entanglement from singlet $|S\rangle_{ij}$ can be obtain from another state $|S\rangle_{il}$ using the permutation operator
$-S^{n}_{m}(i)|S\rangle_{il}=|S\rangle_{ij}$ and $S^{n}_{m}(i)|S\rangle_{ij}=|S\rangle_{ji}$(see fig.1), that reflect the cyclic structure of permutation operator.
The singlet formed by a coupling of sublattice usually in a certain direction (for example, pairwise in Fig.1), and for the triplet it has 
$-S^{n}_{m}(i)|S\rangle_{ik}=0$ and $S^{n}_{m}(i)|S\rangle_{jl}=0$.

 The local magnetization of per plaquette can be written as 
the sum of magnetization in every site,
\begin{equation}   
\begin{aligned}
M=s_{i}\cdot s_{j}+s_{j}\cdot s_{k}+s_{k}\cdot s_{l}+s_{l}\cdot s_{i}=\frac{hc}{\tau e}\frac{\phi}{2\pi}
 \end{aligned}
\end{equation} 
due to the $90^{\circ}$ structure of squattes.
$1/\tau$ is the filling fraction in nesting fermi surface\cite{Li Y Q}and it's a integer constrain essentially, $\phi$ is a virtual twist of the bondary condition
 \cite{Melko R G}which affected deeply by the staggered flux along plaquettes, and it's constrainted by $\sum_{i}\phi_{i}=0 (mod\ 2\pi)$ \cite{Yao H}.

To keep the symmetry form of spin parameter space, we define the spin density wave (SDW) order parameter $\vec{n}$ is a normal vector in the planar configuration in spin space, 
which has $\vec{n}=L^{-d}\langle (\frac{M}{L^{-d}})^{2}\rangle=L^{d}\langle (M)^{2}\rangle$, where $M=L^{-d}\sum_{n}(-1)^{n'}\vec{s}_{n}$ (L, size of square lattice; 
d, dimension).
The $\vec{n}$ is momentum p-dependent and, specially, in the position $(\pi,\pi)$ of the N\'eel phase 
it has $\vec{n}(\pi,\pi)=\frac{N^{2}(N+1)}{12}\langle (\sum_{n}S^{1}_{1}(n))^{2}\rangle$\cite{Harada K}, here the summarize index $n$ stands for four sites in squares,
 i.e., $n=i,j,k,l$.

The thermal averaging of permutation operator corrletion function $\langle c^{\dag}_{iN}c_{iN}\rangle=\frac{1}{\tau}-\frac{\vec{n}}{\tau}+\vec{n}\delta_{n,\mu}$ \cite{Li Y Q},
and it's also the thermodynamical correlation function. In fact, the unbroken order parameter

\begin{equation}   
\begin{aligned}
\vec{n}=L^{-d}\sum_{n}e^{-iqn}\langle s_{n}\ s_{n'}\rangle
\end{aligned}
\end{equation}

where $n$ and $n'$ is site index, and $n'=n,n+x,n+y,n+x+y$. The spin-spin correlation term in equ.(5) structure different spin models which corresponds to different
scaling dimensions. Indeed, the equ.(5) is the Fourier transformation of the correlation function. The spin models in one plaquette can be composed of the SU(2)
Casimir operator. This SDW order parameter suitable if the long range order is well-preserved or the long range order is broken but the SDW states is still stable.

Ref.\cite{Beach K S D} and Ref.\cite{Harada K} both indicate than the staggered magnetization $M$ decreases with the increasing lattice size $L$. Combined with the Ref.\cite{Banerjee A},
the $L$-dependent staggered magnetization have relation of $\langle M^{2}\rangle\propto L^{-d}$ for VBS phase and
 $\langle M^{2}\rangle\propto L^{-(1+\eta)/d}$ for N\'eel phase, where $\eta$ is a N\'eel-dependent parameter. The simulation result of the latter case
of N\'eel state 
have showed in the Fig.4, the fitting line of SU(4) system in two-dimension (d=2) reflect the staggered magnetization obey the power law, the fitting result of 
$\langle M^{2}\rangle=A\cdot L^{-(1+B)/2}$ is $A=2.059\pm 0.18477$ and $B=1.639\pm 0.10282$. According to the data of Fig.4, I count the first- and second-oder Binder
cumulant and its z component one using the formulas from Ref.\cite{Wenzel S,Wang L} 
                   $Q_{1}=\langle M^{2}\rangle/\langle M\rangle ^{2}$, $Q_{2}=\langle M^{4}\rangle/\langle M^{2}\rangle ^{2}$ 
and $Q_{1}^{z}=\langle (M^{z})^{2}\rangle/ \langle M^{z}\rangle ^{2}$， $Q_{2}^{z}=\langle (M^{z})^{4}\rangle/ \langle (M^{z})^{2}\rangle ^{2}$.
The result is reflected in the Fig.5, from the behavior of Binder cumulant, it's show that the second-oder one is bigger that the first-order one, and it shows that
the behavior of $Q$ in critical regime has a comparatively large drop. The critical phenomenon can be extracted in the plot, it show the vanish of long range order.

\begin{figure}[!ht]
   \centering
   \begin{center}
     \includegraphics*[width=0.8\linewidth]{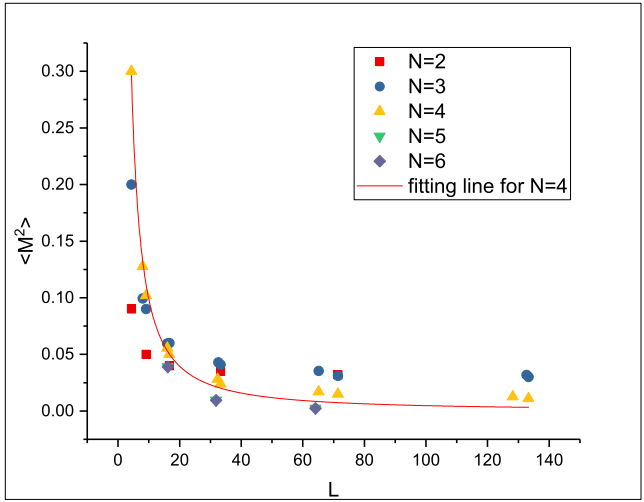}
   \caption{The staggered magnetization $\langle M^{2}\rangle$ as a function of lattice linear size $L$ in N\'eel state of SU(N) system. The red line is the nonlinear 
fitting of value of $\langle M^{2}\rangle$ for $N=4$. The range of various size $L$ is choosen from 0 to 140.}
   \label{fig:user-item-bipartite}
   \end{center}
\end{figure}

\begin{figure}[!ht]
   \centering
   \begin{center}
     \includegraphics*[width=0.8\linewidth]{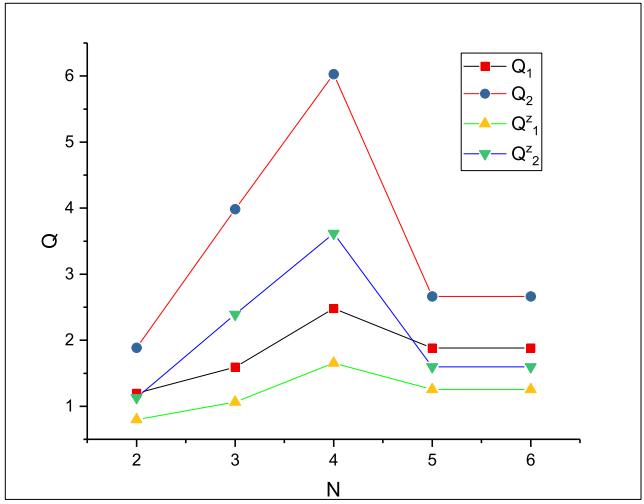}
   \caption{The first- and second-oder Binder cumulant $Q_{1}$ and $Q_{2}$ and the z-component one which calculated from the z-component $\langle (M^{z})^{2}\rangle$ 
versus N. The values of staggered magnetization is take the average of Fig.4 in a size range
which plot in the Fig.4.}
   \label{fig:user-item-bipartite}
   \end{center}
\end{figure}

In this model which both spin and orbit are concerned both, the broken of SU/SO symmetry is accompanied by the occurrence of phase transition (see fig.2).
we define the coupling strength between the singlet and triplet as $J_{3}$, then we can form the Hamiltonian of interaction between two states of singlet and triplet 
which causing by the spin-orbit coupling as:

\begin{equation}   
\begin{aligned}
H_{ST}=\sum_{a,b=1,0,-1,S}J_{eff}|a\rangle \langle b|+J_{2}(|T\rangle \cdot s)+J_{3}(|ST\rangle \cdot s)
\end{aligned}
\end{equation}

Here $J_{eff}$ is the renormalized interaction which consist of the mediate effect between states of singlet and triplet \cite{Ohashi Y}, i.e., the third nearest 
neighbour sublattice participate in the interaction between the two states which studied.
s is the spin operator defined as $s=\frac{1}{\sqrt{2}} \sum c^{\dag}_{i,N}\sigma c_{i,N}$.

Since SU(N) symmetry algebra has $[S^{m}_{n},S^{i}_{j}]=\delta_{mj}S^{i}_{n}-\delta_{ni}S^{m}_{j}$, where $\delta_{mn}=\left\{c_{m},c_{n}^{\dag}\right\}$, so the singlet satisfie
$\sum[S^{m}_{n}(i)-\frac{\delta_{mn}}{N}]|S^{\dag}\rangle=0$ \cite{Coleman P},the $|S^{\dag}\rangle$ contain creation operator only and without the corresponding conjugate one.
 To more clearly illustrate this, we use the Young tableaux for SU(N) system, we konw in the condition of $N\leq 4$, it can
be always description the SU(N) by $N'(N'\leq N)$ sites system. In fig.3,we denote $n_{1}$ and $n_{2}$ as two column of sites,and $n_{col1}+n_{col2}=N$, here we only consider
 the N-site-occupied system.

\begin{figure}[!ht]
   \centering
   \begin{center}
     \includegraphics*[width=0.8\linewidth]{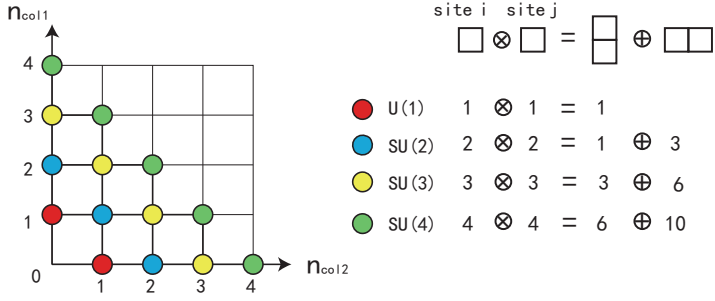}
   \caption{Schematic diagram of Young tableauxs of a two-column model (represente the two orbital state considered here). $n_{col1}$ and $n_{col2}$ are the occupied 
number of sites in two orbits, respectively. that occupied by 
atoms in SU(N) system, and have $n_{col1}+n_{col2}=N$. We only reveal the case of SU(N) system in N-site model, the situation of less sites-occupied can be seen easily.
Every circles in chessboard chart corresponding to different dimensionality. The single-box Young tableaus in the right-side stands for the nearest-neighbor sublattices.}
   \label{fig:user-item-bipartite}
   \end{center}
\end{figure}
 
As a example of SU(N) system ($N=1,2,3,4$)(see fig.3), I use the occupied number of sites in two column of Young tableauxs, and show
the fusion mechanism in the construction of states\cite{Beach K S D}.

Indeed, it's assume for odd number electrons (half-integer spin) per unit site. So considering 
the Fermion form of singlet take N sites and
system-invariance, we have $|S^{\dag}\rangle =\frac{1}{\sqrt{N!}}\sum_{\langle i,j,k,l\cdots\rangle} c_{i,1}^{\dag}c_{j,2}^{\dag}c_{k,3}^{\dag}c_{j,4}^{\dag}\cdots |0\rangle$ 
\cite{Li Y Q,Beach K S D},
here the ellipsis contains N terms due to N-site occupied and $N$ is equal to number of bonds in plaquettes, $|0\rangle$ is the zero flux ground state. It's a framework involve both the spin 
and orbitals degree of freedom which can be 
reflected in equ.(1) and (3). But in more complex square lattice models\cite{Ueda K}, which with broken-square-lattice symmetry but remain the global symmetry.
In this spin-consider-only picture, the N\'eel and VBS ground state are degenerate \cite{Li Y Q} and found to be direct coexistence in the transition point\cite{Banerjee A,Ueda K}.
In critical region, the behavior of different observables is become L-independent as shown in Ref.\cite{Wenzel S} and Ref.\cite{Kaul R K}.

The Hamiltonian in usually Heisenberg model of slave Fermion framework $H=\sum_{\langle i,j \rangle}J_{12}s_{i}s_{j}$, 
the bilinear fermion spin operator $s_{i}=\frac{1}{2}\sum c^{\dag}_{i,N}\sigma c_{i.N}$, where $\sigma$ is the pauli matrix which reflect the overlap of these two 
wave function of two atom state, and $s_{j}$ is as the same form.

Every sites in this N-sites system has a gapped spin liquid since the time and space reversal symmetry \cite{Yao H} prevent the emergence of Weyl point and 
guarantee the gapped. 
This leads to importan propertices of VBS phase, and play an important role in the gapless spin liquid\cite{Wang C}.
The quantum phase transition process in the fig.2 with relatively small spin gap for its left and right sides, and since the closing of spin gap may lead to the
spin-reducing magnon-excitation, so the BEC appears, and bosonics began to cohesion under condensing conditions. A direct result is the Fermions dominates, and furthermore,
the vortices which stable in each sublattice of the superfluid or superconductor region \cite{Machida M} may turn to the rotating spin spirals due to the bosonic condensed. 
 The coupling between these quantum-dimers also become stronger
during the colding process. 
The fermions excitation in the above half-integer spin sites also help to reach the gapless collinear $N'(N'\leq N)$ state.
Addition to the quantum dimer phase, the plaquette resonating-valence-bond(PRVB) phase
may even more dominated than the N\'eel phase in the front part of the phase diagram of fig.2\cite{Ueda K}. It's a paramagnet phase for this RVB state which 
can reached by transition from other quantum magnet phases by sharp change of spin susceptibility. It's relate to critical propertice of spin 
short-range-order instrinsically and has been found that at the frequence of about 4.5 for N\'eel state which showed in the Ref.\cite{Ghaemi P}.

Since it's clear for the neglection of fluctuation in the mean-filed theory, it's a good way for using the perturbation theory to explorate the effect of fluctuation 
to the system, for instance the second order perturbation \cite{Ueda K}or fourth order one\cite{Machida M}.
 Expansively, since the multispin interaction break
the N\'eel order (one of the consequence is the RVBS state), we know that in the AF short range frustrating interaction system, the nearest 
neighbor AF singlet $J_{1}$ 
is help to access the N\'eel, and the nest nearest neighbor coupling $J_{2}$ have a frustrating effect on N\'eel state.
So per bond in one site from
this N-sites system in a square lattices can be described in the plural form $A\pm i\Delta$, where A is the bonds amplitude due to the nonzero flux staggered field 
along plaquettes\cite{Ghaemi P} and $\Delta$ is a energy gap of spinons.

Note that the staggered field in RVB is not always accompanied by a broken long range order when it destory the SU(N) symmetry, the remain long range order guarantee
the stability of the order parameter, which reflected in the stable spin liquid physically. In this case, the eigenfrequency $\omega=0$, so the order parameter 
shows like equ.(5). But in the unstable situation, the eigenfrequency is nonzero and therefore exist a free-diffusive behavior for the disorded order phase,
and the exponent term is become $e^{i(qr-wt)}$.
Usually, since the remain symmetry U(1) has the propertice of gauge flux conservation\cite{Ghaemi P}, the local conserved fluxes has the translation invariance
$[\vec{n}_{n},H]=0$, 

Ref.\cite{Ueda K} puts forward a spin gap of $\Delta=0.11J_{1}$, by the counting of the second order perturbation gap fomular in this artivle, I found that, for the 
RVB state, the spin gap appear in the point of $J_{1}=J'=0.5369$ ($J'$ is the bond between plaquettes), but for dimer state theresult is negative. So it suggest that
the spin gap is belong to the RVB state. In transition region the singlet pair is replaced by a short range and may become a charged
Cooper pairs by the doping which make this RVB state become T-dependent \cite{Baskaran G}. Since the self-consistently of RVB wave function, it can reasonably derive as below.

The bond between site $i$ and $j$ in fermion repretation is 
$H_{ij}=-\sum_{\langle i,j \rangle L}S^{i}_{j}(L)S^{j}_{i}(L)+h.c.$, and the expression of bonds $H_{jk}$, $H_{kl}$ and $H_{li}$
has the same form, where $L$ is the number of bonds, and $S(L)$ here show the average of bond amplitude. For the spin stiffness of this fermions interaction in the sites is $\rho_{s}=L^{d}\partial
^{2}f_{0}/\partial \phi^{2}_{n}$, where N is the total nunmer of bonds, $f_{0}$ is the ground state energy density function which relate to the reduced coupling
factor $(g-g_{c})/g_{c}$, size $L$ and the correlation length of system \cite{Albuquerque A F}, and the site index $n=i,j,k,l$.
From the graphs of Ref.\cite{Wenzel S,Laflorencie N}, it can be directly found that $\rho_{s}$ is inversely proportional to $L$, consider the expression, the $f_{0}$
is inversely proportional to $L$.

    In a plaquette lattice which consist of four bonds (see fig.1b), a complex bounds expression come up as $A+i\Delta$ for horizontal bonds's energyand and
 $A-i\Delta$ for vertical bonds.
Using the above permulation operator $S(L)$ to describe this amplitudes, we have horizatal and vertical bonds expression as $S^{i}_{j}(L)+i\Delta$ and $S^{k}_{l}(L)-i\Delta$
respectively, and the resonal valence bond state can be written as

\begin{equation}   
\begin{aligned}
H=&\ \sum_{\langle i,j \rangle}([S^{i}_{j}(L)+i\Delta]S^{\dag i}_{\ j}(L)+[S^{i}_{j}(L)-i\Delta]S^{\dag j}_{\ i}(L))\\
  &+\sum_{\langle j,k \rangle}([S^{j}_{k}(L)+i\Delta]S^{\dag j}_{\ k}(L)+[S^{j}_{k}(L)-i\Delta]S^{\dag k}_{\ j}(L))\\
  &+\sum_{\langle k,l \rangle}([S^{k}_{l}(L)+i\Delta]S^{\dag k}_{\ l}(L)+[S^{k}_{l}(L)-i\Delta]S^{\dag l}_{\ k}(L))\\
  &+\sum_{\langle l,i \rangle}([S^{l}_{i}(L)+i\Delta]S^{\dag l}_{\ i}(L)+[S^{l}_{i}(L)-i\Delta]S^{\dag i}_{\ l}(L))\\
  &+H_{ij}+H_{jk}+H_{kl}+H_{li}
\end{aligned}
\end{equation}

That involve two important quantities $S(L)$ and $\Delta$, and the author expect to belive they relate to the spin stiffness and fermion (or dirac) nodals,
 respectively. In fermi points, these two qutities equal to zero due to its special position in the momentum space,i,e., become stable and gapless.
 The four spin result had given in the Ref.\cite{Machida M} 
and Ref.\cite{Ghaemi P}, they both contain the spin index and node index, and the author suggest that the node index is connest to the gapless $\mathbb{Z}_{2}$ topology here
which have a feature of chiral degrees of freedom. that's beacause in this fermions interaction ring, the U(1) gauge field is broken by fluctuation and turn into the
 $\mathbb{Z}_{2}$ topology, a result is that the spin-spin correlation in SDW order parameter (Equ.(5)) is domainted totally by the short range and vanish 
identically in long distance\cite{Baskaran G}.
                 
The $S(L)$ here is a description of amplitude which similar to hopping parameter but more suitable for square cells. 
The value of bonds energy can be written as the simple form of Bogoliubov spectrum $E(k)=\sqrt{\varepsilon^{2}(k)+\Delta^{2}(k)}$, where $\varepsilon(k)$ reflect
the behavior of movement in the momentum space and $\Delta(k)$ represent the self-consistent order parameter. Here define two vector in momentum space as
${\bf k_{1}}={\bf k_{x}}sin\phi+{\bf k_{y}}cos\phi$ and ${\bf k_{2}}=-{\bf k_{x}}sin\phi+{\bf k_{y}}cos\phi$, where $\phi$ is the angle of these two vectors.
$\varepsilon(k)$ and $\Delta(k)$ can be written as $\varepsilon(k)=2(S(L)cosk_{1}+S'(L)cosk_{2})$ and $\Delta(k)=2(\Delta sink_{1}+\Delta'sink_{2})$, where $k_{1}$
and $k_{2}$ corresponding to different hopping amplitudes and gaps.
According to the formula of Ref.\cite{Ghaemi P}, the critical coupling ratio is obtained $g_{c}=\frac{1}{L^{2}}\sum_{k}\frac{1}{E(k)}=\frac{1}{L^{2}}\sum_{k}
\frac{-2L^{4}N^{2}\pm\sqrt{\varepsilon^{2}(k)+\Delta^{2}(k)+4L^{8}N^{4}}}{\varepsilon^{2}(k)+\Delta^{2}(k)}$ . Equ.(7) reflect the antiferromagnetic multispin
interaction, the situation which conclude the ferromagnetic one is present in the Appendix.

Summarizing, the bond amplitude under the influence of fluctuation is closely relate to the free-energy of the whole sytem. In this article, the spin liquid is considered
as a intermediate phase in the symmetry-broken SU(N) system, and the analysis of critical phenomenon in the transition region as well as the deduce of system
free-energy is presented.

\begin{appendix}
\section{A brief addition of computation of ferromagnetic and antiferromagnetic spin interaction}
Fig.6 show the simplified model of ferromagnetic interaction in direction b c and antiferromagnetic one in diraction a,
define these three direction as ${\bf n_{a}}=(\frac{\sqrt{2}}{2}e,\frac{\sqrt{2}}{2}e)$, ${\bf n_{b}}=(\frac{\sqrt{2}}{2}e,-\frac{\sqrt{2}}{2}e)$, ${\bf n_{c}}=
(-e,0)$, here $e$ is the length of per horizontal or vertical bond.
The amplitude is $S^{a}(L)$ $S^{b}(L)$ $S^{c}(L)$ respectively. Using the technique put forward in Ref.\cite{Hasegawa Y},
the Hamiltonian is obtained

\begin{equation}   
H=\sum_{k}   [(S^{a}(L)e^{-iq{\bf n_{a}}})+(S^{b}(L)e^{-i  {\bf k_{1}}} )+(S^{c}(L)e^{-i {\bf k_{2}}})]     S^{\dag}(L)+h.c
\end{equation}

where ${\bf k_{1}}=\frac{\sqrt{2}}{2}{\bf k_{x}}+\frac{\sqrt{2}}{2}{\bf k_{y}}$ and ${\bf k_{2}}=\frac{\sqrt{2}}{2}{\bf k_{x}}-\frac{\sqrt{2}}{2}{\bf k_{y}}$.
The energy according to Ref.\cite{Hasegawa Y} is

\begin{equation} 
\begin{aligned}  
E^{2}(k)=&(S^{a}(L))^{2}+(S^{b}(L))^{2}+(S^{c}(L))^{2}+2S^{a}(L)S^{b}(L)cos(e{\bf k_{1}})\\
         &+2S^{b}(L)S^{c}(L)cos(e{\bf k_{2}})+2S^{b}(L)S^{c}(L)cos(e{\bf k_{1}}-{\bf k_{2}})
\end{aligned}
\end{equation}

\begin{figure}[!ht]
   \centering
   \begin{center}
     \includegraphics*[width=0.4\linewidth]{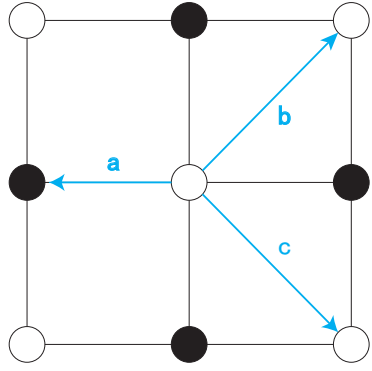}
   \caption{Illustration of simplified model about both the ferromagnetic and antiferromagnetic coupling in a layer planar.}
   \label{fig:user-item-bipartite}
   \end{center}
\end{figure}

\end{appendix}

\end{large}
\renewcommand\refname{References}

\end{document}